\documentstyle[12pt]{article}
\newcommand{\be}{\begin{equation}}
\newcommand{\ee}{\end{equation}}
\newcommand{\bea}{\begin{eqnarray}}
\newcommand{\eea}{\end{eqnarray}}
\newcommand{\bi}{\begin{itemize}}
\newcommand{\ei}{\end{itemize}}
\newcommand{\bc}{\begin{center}}
\newcommand{\ec}{\end{center}}
\newcommand{\bfl}{\begin{flushleft}}
\newcommand{\efl}{\end{flushleft}}
\newcommand{\bfr}{\begin{flushright}}
\newcommand{\efr}{\end{flushright}}
\newcommand{\f}{\frac}

\newcommand{\au}{\u{a}}

\def\6{\partial}  
\def\g{\gamma} \def\d{\delta}

 \def\k{\kappa} \def\l{\lambda}
\def\m{\mu}  \def\x{\xi} 
\def\r{\rho}  
 \def\ph{\varphi} 
\def\o{\omega}  \def\D{\Delta}
  
\def\PH{\Phi}

\def\non{\nonumber\\}
\def\={\!\!\!&=&\!\!\!}
\def\+{\!\!\!&&\!\!\!+~}
\def\-{\!\!\!&&\!\!\!-~}

\def\cb{\bar{c}}
\def\Cb{\bar{C}}
\def\pb{\bar{\partial}}
\def\mb{\bar{\mu}}
\def\zb{\bar{z}}

\def\lb{\bar{\lambda}}
\def\phij{\phi_{j,\bar{j}}}
\def\phiju{\phi_{j_1,\bar{j}_1}^{p_1,q_1}}
\def\phijn{\phi_{j_n,\bar{j}_n}^{p_n,q_n}}
\def\PHj{\Phi_{j,\bar{j}}}
\def\PHjp{\Phi_{j,\bar{j}}^{p,q}}
\def\PHjpu{\Phi_{j_1,\bar{j}_1}^{p_1,q_1}}
\def\PHjpn{\Phi_{j_n,\bar{j}_n}^{p_n,q_n}}
\def\PHkru{\Phi_{k_1,\bar{k}_1}^{r_1,s_1}}
\def\PHkrm{\Phi_{k_m,\bar{k}_m}^{r_m,s_m}}
\def\bX{{\bf X}}
\newcommand{\equ}[1]{(\ref{#1})}

\renewcommand{\AA}{{\cal A}}

\newcommand{\CC}{{\cal C}}

\newcommand{\MM}{{\cal M}}

\newcommand{\WW}{{\cal W}}

\newcommand{\journal}[4]{{\em #1~}#2\,(19#3)\,#4;}

\newcommand{\pr}{\journal {Phys. Rev.}}

\newcommand{\jmp}{\journal {J. Math. Phys.}}

\newcommand{\np}{\journal {Nucl. Phys.}}
\newcommand{\pl}{\journal {Phys. Lett.}}

\newcommand{\annp}{\journal {Ann. Phys. (N.Y.)}}

\begin{document}
\title{BRST COHOMOLOGY IN BELTRAMI PARAMETRIZATION}
\author{{\em Liviu T\u{a}taru and Ion V. Vancea,}  \\
Department of Theoretical
Physics \\
Babes - Bolyai University of Cluj, 3400 Cluj-Napoca \\
Romania}
\date{1 April 1995}
\maketitle
\begin{abstract}
We study the BRST cohomology within a local
conformal Lagrangian field theory model built on a two dimensional
Riemann surface with no boundary. We deal with the case of the
complex structure parametrized by Beltrami differential and the
scalar matter fields. The computation of {\em all} elements of
the BRST cohomology is given.
\end{abstract}
\setcounter{page}{0}
\thispagestyle{empty}

\newpage
\section{Introduction}
The Beltrami differentials \cite{bers} have turned out
to be basic variables for parametrizing complex structures for a
bidimensional theory, in a conformally invariant way \cite{b,st,bb}.
The advantage of
the Beltrami differentials is the fact that they changes only under
reparametrization transformations and diffeomorphisms while the
Weyl rescaling is factorized out. Thus the Weyl degree of freedom
is eliminated from the very beginning and some advantages in the
quantization are reached.

In this paper we shall describe models for which a conformal
matter field
of
weight $(j,\bar{j})$ coupled to a 2D gravity, is characterized by the
Beltrami
differentials, in a reparametrization invariant way
\cite{b,s,bbg1,bbg2,g}. To accomplish
this aim we shall introduce a BRST symmetry \cite{st,d,b,brst}
carried out
through a nilpotent operator $s$ and we shall calculate its
cohomology
group $H(s)$ within a {\em local} Lagrangian field theory
formulation. We
are not going to characterize these models by a specific
conformally
invariant classical action but rather we shall specify the field content
and the gauge invariances of the classical theory. In this
framework
the search for the invariant Lagrangians, the anomalies and the
Schwinger
terms can be done in a purely algebraic way, along the line of
algebraic
topology \cite{d,btp,band}. In fact, the main purpose of our search
is to
find out {\em all} nontrivial solutions of the equation
\be
s\AA=0 \label{con1}
\ee
with $s$ the nilpotent BRST differential and $\AA$ is an integrated
local
functional $\AA=\int d^2 x f.$  The condition \equ{con1} can be
translated
into the local descent equations \cite{st,wss,s}
\bea
s\o_2+d\o_1=0 \hspace{1cm},\hspace{1cm}s\o_1+d\o_0=0 \non
s\o_0=0. \label{desc}
\eea
where $\o_2$ is a 2-form with $\AA=\int \o_2 $ and $\o_1,\o_0$
are {\em local} 1- and 0-forms, respectively. It is well known \cite{brand}
that the descent equations \equ{desc} end, for the
Beltrami parametrization, always with a nontrivial 0-form $\o_0$ and
that their "integration" is trivial
\be
\o_1=\d\o_0\hspace{1cm},\hspace{1cm}\o_2=\f{1}{2}\d^2\o_0,
\ee
where the operator $\d$ was introduced by Sorella \cite{st,s} and
it allows to express the exterior derivative $d$ as a BRST
commutator:
\be
d=-[s,\d]. \label{sorella}
\ee
Thus it is sufficient to find out the general solution of
the equation
\be
s\o_0=0, \label{basic}
\ee
in the space of local functions of the fields and their derivatives
i.e.
to calculate the BRST cobomology group $H(s)$.

In this paper we shall calculate {\em all } elements of $H(s)$ for
the
string theory in the Beltrami parametrization
in the presence of one scalar matter field of weight $(0,0)$. The basic
ingredients of our
calculations are the choice of an appropriate new basis and the
existence in
this basis a contracting homotopy, which reduces considerably
the number of the elements from the
basis for
the solutions of \equ{basic}. In this way we shall obtain a very
limited
possible solutions of \equ{basic} which can be listed and studied.
We want to
stress that the basis used in this paper is very closed to the one
proposed  by Brandt, Troost and Van Proyen, in a very interesting
paper
\cite{btp} but the BRST transformations of this basis and
the contracting homotopy differ and our method can be easily
generalized for
other models as superstring model in the super-Beltrami
parametrization
\cite{tv} and $W_3$-gravity \cite{tb}.

The paper is organized as follows. In Sect.2 we briefly recall the
Beltrami
parametrization and
its BRST  symmetry. In Sect. 3 we define the differential algebra
of
all fields and their derivatives $\AA$ and we show that it can be
split in
a contractive part $\CC$ and a minimal one $\MM$. Only the minimal
part
does contribute to the BRST cohomology. In Sect. 4 we introduce a
new
basis and show that, in the presence of the nonlocal fields $\ln \l $
and $ \ln \bar{\l} $
the equation \equ{basic} has nontrivial
solutions in a
very small subalgebra, which we are going to describe. In this
subalgebra we find {\em all} nontrivial elements of $H(s)$. Thus we
can find out all
solutions of the descent equations \equ{desc} In Sec. 5 the
cohomology group $H(s)$ and the
solutions of the decent equation \equ{desc} are constructed for theories
{\em without} the fields $\ln \l $ and $ \ln \bar{l} $.

\section{The diffeomorphism BRST cohomology}
\setcounter{equation}{0}
Let we start by introducing the setup for the string theory in the
Beltrami parametrization. We will work on a Riemann surface M
equipped
with a complex structure or, equivalently, with a conformal class
of
metrics \cite{bers}. Using the complex notations $dz=dx+idy,
d\bar{z}=
dx-idy$ the line element associated to the metric can be written
as:
\be
ds^2=\mid \r \mid^2\mid dz+\m d\bar{z} \mid^2
\label{bars}
\ee
where $\r$ and $\m$ are smooth complex-valued functions of
$z,\bar{z}$
and the positive-definiteness of the metric is expressed by the
condition
$\mid\m\mid <1$. The function $\r$ is usually called the
{\em conformal factor} and $\m$  {\em the Beltrami differential
(or parameter)} \cite{bers} and \equ{bars} is often called the
{\em Beltrami parametrization} of the metric. \cite{bb}. The line
element $ds^2$ can be written in terms of isothermal coordinated
$(Z,\bar{Z})$ such that $ds^2\sim\mid dZ\mid^2$. These isothermal
coordinates are defined by
\be
dZ=\l(z,\bar{z})\left[dz+\m d\bar{z}\right]
\ee
with $\l$ a smooth complex-valued function, called {\em the
integrating factor}. The condition $d^2=0$ yields
\be
(\bar{\6}-\m\6)(\ln \l)=\6\m .\label{diffeq}
\ee
The line element $ds^2$ has a very simple form in the isothermal
coordinates
\be
ds^2=\mid dZ \mid^2.
\ee
Despite of the tact that the conformal factor $\r$ and the
integrating
factor $\l$ look  very similar, they have different transformations
laws and they are very different in many respects.

The matter fields, in our models, are realized by local tensor
fields
\be
\PH_{j,\bar{j}}(Z,\bar{Z})dZ^jd\bar{Z}^{\bar{j}}\label{matter}
\ee
of the weight $(j,\bar{j})$ invariant under change of holomorphic
charts. The matter fields $\ph_{j,\bar{j}}$ are defined by
\equ{matter} written in terms of the local coordinates
$(z,\bar{z})$:
\be
\PH_{j,\bar{j}}(Z,\bar{Z})dZ^jd\bar{Z}^{\bar{j}}=
\ph_{j,\bar{j}}(z,\bar{z})(dz+\m
d\bar{z})^j(d\bar{z}+\bar{\m}dz)^{\bar{j}},
\ee
with
\be
\PH_{j,\bar{j}}=\f
{\ph_{j,\bar{j}}(z,\bar{z})}{\l^j(z,\bar{z})\bar{\l}^{\bar{j}}(z,
\bar{z})}.
\ee
The BRST symmetry can be obtained by considering an infinitesimal
change of the coordinate $(z,\bar{z})$ generated by a vector field:
\be
\x.\6=\xi\6_z+\bar{\x}\6_{\bar{z}}=\x\6+\bar{\x}\bar{\6},
\ee
and then replacing the parameters $(\x,\bar{\x})$ by the ghosts
$(c,\bar{c})$.
Thus the BRST differential $s$ acts on $Z$ and
$\ph_{j,\bar{j}}$ as the Lie derivatives
\bea
sZ&=&L_{c\cdot\6}Z=\l (c+\m\bar{c}) \label{zet} \\
s\ph_{j,\bar{j}}&=&L_{c\cdot\6}\ph_{j,\bar{j}}=(c\cdot\6)\ph_{j\bar{j}}
+\\
&+&[j(\6 c+\m \6\bar{c})+\bar{j}(\bar{\6}\bar{c}+\bar{\m}
\bar{\6}c)]\ph_{j\bar{j}}.\label{phi}
\eea
The operator $s$ acts as an antiderivation from the left and the
{\em graduation} is given by adding the form degree to the ghost
number.

The corresponding transformation laws of $\m$ and $\l$ follow
by evaluating the
variation of $dZ$ in two different ways
$$
s(dZ)=-d(sZ)= -d[\l (c+\m\cb )]
$$
and
$$s(dZ)=s[\l (dz+\m d\bar{z})]
$$
By comparing the different coefficients of $dz$ and $d\bar{z}$ one
finds
\bea
s\m&=&(c\cdot\6)\m -\m (\6c+\m
\6\bar{c})+\bar{\6}c+\m\bar{\6}\bar{c}
\label{miu}\\
s\l&=&\6[\l(c+\m \bar{c})].\label{lambda}
\eea
The nilpotency of $s$ requires
$$
0=s^2Z=[sc-c\6c]\l $$
and thereby
\be
sc=c\6 c. \label{ce}
\ee
It is very convenient to replace the ghosts
$(c,\bar{c})$ with the Becchi's reparametrization \cite{b}
\be
C=c+\m \bar{c}\hspace{1cm},\hspace{1cm}\bar{C}=\bar{c}+\bar{\m}c.
\label{newghosts}
\ee
This reparametrization ensures the {\em holomorphic factorization}
of the
BRST variations of $\m$ and $\l$. \cite{brst,bb}.
Eqs.\equ{zet}-\equ{ce}
can be rewritten as
\bea
sZ&=&\l C \label{brst1}\\
s\ph_{j\bar{j}}&=&CD\ph_{j\bar{j}}+\bar{C}\bar{D}\ph_{j\bar{j}}+\non
&+&[j(\6C)+\bar{j}(\bar{\6}\bar{C})]\ph_{j\bar{j}} \label{brst2}\\
s(\ln\l)&=&\6C\label{brst3} \\
s\m&=&\bar{\6}C+C\6\m-\m\6C \label{brst4}\\
sC&=&C\6C \6(\ln \l )C \label{brst5}
\eea
and their complex conjugate expressions,with
\be D=\l\f{\6}{\6 Z}=\f{1}{1-\m\bar{\m}}(\6-\bar{\m}\bar{\6})
\hspace{1.3cm}
\bar{D}=\bar{\l}\f{\6}{\6\bar{ Z}}=\f{1}{1-\m\bar{\m}}(\bar{\6}
-\m\6)
\ee
are the "covariant dertivatives" \cite{bbs,wss,bl1,bl2}.
Now we introduce the ghost number (or Faddeev-Popov charge)
gh=g,which is
{\em one} for the ghost fields $c$ and $\cb$ (or equivalently $C$
and $\Cb$ ) and the BRST differential $s$ and zero for the other
fields $\{ Z, \bar{Z}, \ph_{j\bar{j}}, \l, \bar{\l}, \m, \mb\}$ and
the differential d.

It is easy to see that the nilpotency of $s$ is equivalent to the
differential
equation \equ{diffeq} and the commutation relations
\be
[s , \6]=[s , \pb]=0 \ee
or equivalently
\be
\{s , d\}=0
\ee
with $$ d=dz\6+d\bar{z}\pb $$
the exterior derivative.

The main purpose of the present paper is to give the most general
solution of the equation \cite{wz,sto}
\be
\label{inteq} s\AA^p=0 \hspace{0.5cm} \mbox{with}
\hspace{0.5cm} \AA^p=\int \D^p_r(z,\bar{z})
\ee
where $\AA^p$ has the ghost number two and $\D^p_r(z, \bar{z})$ is
a
{\em r-form} with the ghost number {\em p}. In eq. \equ{inteq} {\em
r} can
take two values $r=1 , 2$ and in these cases we have the  1-form
descent equations \cite{witten,wz,bl2} and two-form descent equations
\cite{bl1,wss}.

In terms of local quantum eq.\equ{inteq} is expressed by the
$s$-cohomology
modulo $d$:
\bea
\label{des1}
s\D^p_1(z,\zb)+d\D^{p+1}_0(z,\zb)=0 \non
s\D^{p+1}_0(z,\zb)=0
\eea
or
\bea \label{des2}
s\D^p_2(z,\zb)+d\D^{p+1}_1(z,\zb)=0 \non
s\D^p_1(z,\zb)+d\D^{p+1}_0(z,\zb)=0 \non
s\D^p_0(z,\zb) =0
\eea

The ladder \equ{des1} or \equ{des2} could be solved thanks to an
operator $\d$
introduced by Sorella  for the Yang-Mills BRST cohomology
\cite{s,st},
bosonic string \cite{wss} and superbosonic string \cite{bss} (see
also \cite{tb} for $W_3$-gravity). The operator $\d$ allows us to
express
the exterior derivative $d$ as a BRST commutator, i.e. :
\be \label{sorella3}
d=-[s , \d]. \ee
 Now it is easy to see that, once the decomposition \equ{sorella3}
has been found, repeated application of the operator $\d$ on the
local
functions  $\{\D^{p+2}_0(z ,\zb) , \D^{p+1}_0(z ,\zb)\}$ that
solve the last equation of \equ{des1} or \equ{des2} given an
explicit
and nontrivial solution for the other cocycles $\D^{p+n}_n(z
,\zb)$.
In other words, with the operator $\d$ we can go from the
cohomology $H(s)$
 to the relative cohomology $H(s~\mbox{mod}~ d)$.

In our theory the operator $\d$ from the docomposition
\equ{sorella} can be defined by
\bea \label{defdel}
\d C&=&dz+\m d\zb \non
\d\Cb &=&d\zb +\bar{\m}dz \non
\d\PH&=&0~~~ \mbox{for}~~~ \PH=\{ \m, \bar{\m}, \l,\lb, \ph_{j,\bar{j}}\}.
\label{delta}
\eea

Now it is ease to verify that $\d$ is of degree 0 and obeys the
following
algebraic relations:
\be \label{sorella2}
d=-[s , d]\hspace{1cm},\hspace{1cm} [d ,\d ]=0.
\ee

To solve the towers \equ{des1} or \equ{des2} we shall
make use of
the following identity
\be \label{tatar}
e^\d s=(s+d)e^\d \ee
that is a direct consequence of \equ{sorella} and \equ{sorella2}
(see \cite{s}). Therefore, once a non-trivial solution. In this
way we
get
\be
(s+d)\left[e^\d\D^{p+n}_0\right]=0\hspace{1cm}(n=1,2).
\label{solut}
\ee
But, as one can see from \equ{defdel}, the operator $\d$ acts as a
translation on the ghosts $(C , \Cb)$
$$
C\rightarrow C+dz+\m d\zb \hspace{2cm}\Cb\rightarrow
\Cb+d\zb+\bar{\m} dz $$
and eq. \equ{solut} can be rewritten as
\be
(s+d)\D^{p+n}_0(C+dz+\m d\zb , \Cb +d\zb +\mb dz, \m, \mb, \l, \lb,
\ph_{j,\bar{j}} )=0.
\label{solut1}
\ee

Thus the expansion of the zero form cocycle $\D^{p+n}_0$ in power
of the one-forms $(dz +\m d\zb , d\zb +\mb dz )$ yields all the
cocycles $ \D_r^{p+n-r}$.
The operator $\d$, defined in \equ{defdel} is closed connected with
the
operators $(\WW, \bar{\WW})$ introduced  in \cite{wss} and defined
as:
\be
\WW =\int dz d\zb \left(\mb\f{\d}{\d \Cb}+\f{\d}{\d C}\right) \ee
\be
\bar{\WW} =\int dz d\zb \left(\m\f{\d}{\d C}+\f{\d}{\d \Cb}\right)
\ee

Our $\d$ is  related to$(\WW ,\bar{\WW})$ by the relation
\be
\d=dz\WW + d\zb \bar{\WW} \ee
and the relations \equ{sorella2} imply
\bea
\{s ,\WW\}=\6\hspace{2cm}\{s , \bar{\WW}\}=\bar{\6}\non
\{\WW ,\WW\}=\{\WW ,\bar{\WW}\}=\{\bar{\WW} ,\bar{\WW}\}=0.
\eea

\section{BRST symmetry in a new basis}
\setcounter{equation}{0}
In his section we are going to solve the equation
\be
s\o_0=0 \label{fund}
\ee
in the algebra of local analytic function of all fields and their
derivatives
$\AA$. A basis of this algebra can be chosen to be
\be
\{\6^p\bar{\6}^q\Psi , \6^p\bar{\6}^q C, \6^p\bar{\6}^q \bar{C} \}
\label{basis} \ee
where $\Psi=\{\m, \bar{\m}, \l, \bar{\l}, \phij \}$ and
$p,q=0,1,2,\cdots$.
However, the BRST  transformations of the elements of this basis
are quite complicated
and there are many terms which can be eliminated in $H(s)$. In fact
the algebra $\AA$ with the BRST differential $s$ form a
free differential algebra, which can be decomposed, by using a
theorem due to
Sullivan \cite{sull}, as a tensor product of a {\em  contractible
algebra} $\CC$ and a {\em minimal} one $\MM$. A contractible
differential
algebra  $\CC$ is an algebra isomorphic to a
tensor product of algebras of the form $\wedge(x ,sx)$ and a
minimal one
$\MM$ is an algebra for which $s\MM\subseteq\MM^+\cdot\MM^+$.
with $\MM^+$ the part  of $\MM$ in positive degree. The remarkable
point in Sullivan decomposition is the fact that the  contractible
part of the
algebra $\AA$ does not contribute to the cohomology grout $H(s)$.
Thus, to calculate $H(s)$ it is enough to separate from the
differential
algebra $\AA$ its minimal part $\CC$ and to calculate the
cohomology group
of $\MM$. Indeed, according to the K\"{u}neth
theorem
$$
H(\MM\otimes\CC)=H(\MM)\otimes H(\CC) $$
since $\CC$ is contractible and its cohomology group is zero.

The separation of the algebra $\AA$ in two parts is easier to be
accomplished
if we introduce  a new basis of variables substituting the fields
and their
 derivatives \equ{basis}. The hew basis consist of
\def\phijp{\phi_{j,\bar{j}}^{p,q}}
\begin{enumerate}
\item the variables $\phijp$ , substituting one-by-one the partial
derivatives $\6^p\pb^q\phij $ of the matter fields
\be
\phijp=\D^p\bar{\D}^q\phij \label{basis1}
\ee
where the {\em even} differentials $\{ \D , \bar{\D} \}$ are
defined by
\be
\D=\left\{ s , \f{\6}{\6 C} \right\} \hspace{1.5cm}
\bar{\D}=\left\{ s , \f{\6}{\6 \bar{C}} \right\}; \label{oper1}
\ee
\item
the ghost variables
\be
C^n=\f1{(n+1)!}\D^{n+1} C\hspace{1.5cm}
\bar{C}^n=\f1{(n+1)!}\bar{\D}^{n+1}\bar{C}. \label{basis2}\ee
\end{enumerate}

All the variables \equ{basis1} and \equ{basis2} have two remarkable
properties. First, they have very simple BRST  transformation
properties, being the basis for the minimal part of $\AA$ and
second,
they have a special total weight which allow us to select only very
few possibilities for the solutions of equation \equ{fund}. The
transformation properties of these variables can be obtained from their
definitions, the commutation relations
\be
\D s=s \D \hspace{1cm},\hspace{1cm}\bar{\D} s=s \bar{\D}
\label{comm}
\ee
and the fact that $\{ \D ,\bar{\D}\}$ are even differentials i.e.
they
satisfy the Leibniz rule
\be
\label{leibniz1}
\D(a b)=(\D a)b  = a(\D b)
\ee
and its generalized form
\be
\label{leibniz2}
\D^n(a b)=\sum_{k=0}^{n}\left (\begin{array}{c} n\\k
\end{array}\right)
(\D^k a)(\D^{n-k}b) .
\ee
The BRST transformation of the basis \equ{basis2} can be witten as
\be
sC=\f{1}{(n+1)!}\D^{n+1}(C\6C)=\sum _{k=-1}^{n-1}(n-k)C^k C^{n-k}=
\f1{2}f_{pq}~^n C^p C^q, \label{brst6}
\ee
where
\be
f_{pq}~^n=(p-q)\d^n_{p+q} \label{brst7} \ee
since the BRST transformaton of C is given by \equ{brst5}, and of course
the complex conjugate expressions.

For the elements of the basis \equ{basis1} the BRST transformation can be
obtained from their  definition and the transformation of
the matter fields $\phij$ \equ{brst2}. Therefore eqs. \equ{leibniz1} and
\equ{leibniz2} yield
\bea
s\phijp =\D^p\bar{\D}^q (s\phij)=\sum_{k=-1}^{p-1}A^k_p C^k\phij^{p-k,q} +
\sum_{k=-1}^{q-1}A^k_q \Cb^k\phij^{p,q-k}=\non
=\sum _{k=-1}^\infty (C^k L_k +\Cb^k \bar{L}_k )\phijp
\label{brst8}
\eea
where
\bea L_k\phij^{p,q}=A^k_p(j)\phij^{p-k,q}\non
\bar{L}_k\phij^{p,q}=A^k_p(\bar{j})\phij^{p-k,q} \label{vir1}
\eea
and
\be
A_p^k(j)=\f{p!}{(p-k)!}[j(k+1)+p-k]. \label{comb}
\ee

On the basis $\{\phijp\}$ the operators $\{ L_k ,\bar{L}_k \}$ represent two
copies of the Virasoro algebra, fact that can be seen from their definitions
\equ{vir1}. This fact was pointed out by Brandt, Troost and
Van Proeyen \cite{btp} for 2D conformal gravity  in a basis formed by
$\phi_{0,0}$ matter fields. In fact the definitions \equ{vir1}
yield
\be
[L_m , L_n]=f_{mn}^k L_k ~~~,~~~[\bar{L}_m ,\bar{L}_n]=
f_{mn}^k \bar{L}_k ~~~,~~~
[L_m ,\bar{L}_n]=0 \label{vir2}
\ee
where $f_{mn}~^k$ are the structure coinstants of the
Virasoro algebra given by \equ{brst7}.

On the other hand, the relation \equ{brst8} shows that {\em on the
basis} $\{\phijp\}$ the generators of the Virasoro algebra
$\{L_k ,\bar{L}_k\}$ have the following expresions:
\be
L_k=\left\{ s ,\f{\6}{\6 C^k}\right\}~~~,~~~\bar{L}_k=\left\{ s ,
\f{\6}{\6 \bar{C}^k}\right\},~~~k>-2 .
\label{vir3}
\ee

Hitherto we have not said anything about the other members of the basis
for the algebra $\AA$ \equ{basis}, i.e. about $\{\m, \mb , \l , \bar{\l}\}$
and their derivatives. The structure of the free differential algebra
$\AA$ strongly depends on the fact that we allow the variables
$\{\ln\l , \ln\bar{\l} \}$ in it.

\section{BRST cohomology with the variable $\ln{\l}$}
\setcounter{equation}{0}

Now if we consider the $\ln\l$ and $\ln\bar{\l}$ as variables in our
differential algebra $\AA$ then the BRST cohomology  has a very simple form.
Indeed, in this case the BRST transformations
\bea
s(\ln\l)=\6 C~~~~+\6(\ln\l)C~~~&,&~~~s(\ln\bar{\l})=\6 \Cb~~~~+\bar{\6}
(\ln\bar{\l})\Cb \non
s\m=\bar{\6}C+C\6\m-\m\6 C~~~&,&~~~s\mb=\6\Cb+\Cb\bar{\6}\mb-\mb\6\Cb
\eea
show that the subalgebra $\CC$ could be generated by the elements
\be
\CC= \{\6^p\bar{\6}^q\m~,~ \6^p\bar{\6}^q\mb~,~
\6^p\bar{\6}^q\l~,~\6^p\bar{\6}^q\bar{\l}~,~ s(\6^p\bar{\6}^q\m)~,~
s(\6^p\bar{\6}^q\mb)~,~s(\6^p\bar{\6}^q\l)~,~s(\6^p\bar{\6}^q\bar{\l}) \}.
\label{contractiv}\ee
A possible candidate for the minimal subalgebra $\MM$ might be
generated by the elements
\be
\MM '=\{C , \Cb , \phijp\} \label{minimal}.
\ee
For example {\em all} the derivatives of $C$ and
$\Cb$ can be expressed as polynomials of the elements of the basis
\equ{contractiv} and \equ{minimal}.

However, as it can be seen by a simple inspection of the BRST transformations
of $\phijp$, the algebra $\MM'$ does not satisfy the condition for a minimal
algebra $$
s\MM'\subseteq \MM'^+\cdot \MM'^+. $$ But we can slightly modify
the subalgebra
$\MM'$ to obtain a minimal one. Instead of the matter fields $\phij$ we
shall use the matter fields $\PHj$ in the $(Z ,\bar{Z})$ complex analytic
coordinates defined by
\be
\PHj(Z,\bar{Z})dZ^j d\bar{Z}^{\bar{j}}=\phij (z,\bar{z})(dz+\m d\bar{z})^j
(d\bar{z}+\bar{\m}dz)^{\bar{j}},
\ee
or
\be
\PHj(Z,\bar{Z})=\f{\phij(z,\bar{z})}{\l^j(z,\bar{z})\bar{\l}^{\bar{j}}(z,
\bar{z})}
\ee
Here it is crucial to point out that the fields $\PHj$ behaves like
{\em scalar quantities} in the $(Z ,\bar{Z})$ coordinates whereas
the old matter  fields $\phij$ have a {\em tensorial} nature in the
$(z , \bar{z})$ coordinates since the diffeomorphism action is performed
in the background coordinates $(z , \bar{z})$.

Since the diffeomorphisms only move the coordinates $(Z,\bar{Z})$,
the BRST transformations for the new fields have the form:
\bea
sZ=\g =\l(c+\m\bar{c})=\l C~~~&,&~~~s\bar{Z}=\bar{\g}=\bar{\l}(\bar{c}+
\bar{\m} c)=\bar{\l}\bar{C} \non
s\PHj&=&(\g \6_Z + \bar{\g} \6_{\bar{Z}})\PHj \non
s\g&=&s\bar{\g}=0  \label{newbrst}
\eea
where the Cauchy-Riemann operators $\6_Z$ and $\6_{\bar{Z}}$ read in
term of the $(z,\bar{z})$ coordinates
\be
\6_Z=\f{\6-\bar{\m}\bar{\6}}{\l(1-\m\bar{\m})}~~~,~~~
\6_{\bar{Z}}=\f{\bar{\6}-\m\6}{\l(1-\m\bar{\m})}.
\ee

The construction presented in the previous section can be accomodated for
these new variables. Indeed one can construct a suitable basis by
introducing the differential operators $(\tilde{\D}~,~\tilde{\bar{\D}})$
defined by
\be
\tilde{\D}=\left\{ s~,~\f{\6}{\6\g}\right\} ~~,~~\tilde{\bar{\D}}=
\left\{ s~,~\f{\6}{\6\bar{\g}}\right\} \ee
and defining
\be
\PHjp=\tilde{\D}^p\tilde{\bar{\D}}^q\PHj \ee
By using \equ{newbrst} one can see that the BRST transformations of
$\PHjp$ have the form
\be
s\PHjp=\g \PHj^{p+1,q} ~+~\bar{\g} \PHj^{p,q+1}=(\g \tilde{\D}+
\bar{\g}\tilde{\bar{\D}})\PHjp.\label{phi1} \ee
The fields $\PHjp$  have, in fact, a very simple form in the $(Z~,~\bar{Z})$
coordinates. Indeed for an arbitrary function of $(Z~,~\bar{Z})$ one can
write
\be
sF(Z~,~\bar{Z})=(\g\f{\6}{\6 Z}~+~\bar{\g}\f{\6}{\6 \bar{Z}})F(Z~,~\bar{Z})
\label{identity} \ee
which allows us to rewrite  $\PHjp$ in a simpler form.
The BRST transformations of $\g,\bar{\g}$ and $\PHj$ \equ{newbrst}
and the identity \equ{identity} yield
\be
\tilde{\D}^2\PHj =\f{\6}{\6\g}\left[s(\6_{Z}\PHj)\right]=\6_{Z}^2\PHj.
\label{newbasis2}
\ee
The relation \equ{newbasis2} can be easily generalized and we eventually
get
\be
\PHjp =\6_{Z}^p\6_{\bar{Z}}^q \PHj .\ee
Therefore the basis
\be
\MM=\{\PHjp~~,~~\g~~,~~\bar{\g}\}~~~~~(p,q=0,1,\cdots) \label{minimal2}
\ee
represents a basis for the minimal subalgebra $\MM$ of the algebra $\AA$.

By using the BRST transformations of $\g$ and $\PHjp$  it is easy
to see that the weights of both $\g$ and $\PHjp$,
 defined in the previous section,  are zero i.e.
\be
L_0\g=L_0\bar{\g}=L_0\PHjp=0~~~~,~~~~\bar{L}_0\g=\bar{L}_0\bar{\g}=
 \bar{L}_0\PHjp=0. \ee
The BRST cohomology group here have to be calculate
only in the basis \equ{minimal2} and in this new basis,
due to the nilpotency of $\g$ and $\bar{\g}$ we have only several
of candidates for the solutions of equation \equ{fund}.
In fact we have
only two possibilities:
\bea
\o_0^{(1)}=c_1\g \PHjpu\cdots\PHjpn~+~
c_2 \bar{\g} \PHkru\cdots\PHkrm=\g \Pi_1~+~\bar{\g}\Pi_2\label{poss1}\\
\o_0^{(2)}=c_3\g\bar{\g} \PHjpu\cdots\PHjpn=\g\bar{\g}\Pi_3. \label{poss2}
\eea
The possibility $\o_0^{(1})$ can be a solution of \equ{fund} only for some
paticulare values of $\Pi_1$ and $\Pi_2$ . Indeed if one use \equ{newbrst}
and \equ{phi1} then we can write
$$
s\o_0^{(1)}=-\g  (\g\6_{Z}+\bar{\g}\6_{\bar{Z}})\Pi_1  -
\bar{\g}( \g\6_{Z}+\bar{\g} \6_{\bar{Z}})\Pi_2 =
-\g\bar{\g}[\6_{\bar{Z}}\Pi_1-\6_Z\Pi_2 ] = 0.
$$
Thus a solution of \equ{fund} in the basis considered in this
section has the form \equ{poss1} with
\be
\6_{\bar{Z}}\Pi_1=\6_Z\Pi_2. \label{poss12}
\ee
The solution of \equ{poss12} has the form
$$
\Pi_1=\6\Pi \hspace{2cm}\Pi_2=\6_{\bar{Z}}\Pi $$ and $\o_0^{(1)}$ is
s-exact
\be
\o_0^{(1)}=(\g \6_Z +\bar{\g}\6_{\bar{Z}})\Pi =s\Pi .\ee
The candidate $\o_0^{(2)}$ is a solution of \equ{fund}
fact that can be seen easily
$$
s\o_0^{(3)}=\g \bar{\g}\left[(\g \tilde{\D}+
\bar{\g}\tilde{\bar{\D}}\right]\Pi=0. $$
We can resume all these disscutions by saying that
in the differential algebra $\AA$, which includes the fields
$\ln\l~,~\ln\bar{\l}$,  the general solution of the equation \equ{fund} is
given by
\be \o_r^p(z,\bar{z})=C(z,\bar{z})\Cb(z,\bar{z})\Pi_r^{p-2}(z,\bar{z})
\d^2_0
+s\o_r^{p-1} \label{generalsol}
\ee
where the redefined ghost fields $C$ and $\Cb$ (see \equ{newghosts})
occur and $\o_r^p$ is a r-form with the ghost p.

These results represent in fact the main results obtained
by Bandelloni and Lazzarini in \cite{bl1,bl2} by using the spectral
sequence method to calculate the local BRST cohomology
modulo $d$. In the present paper we have obtained these crucial results
just by using a very convenient basis and Sullivan's theorem which
has allowed us to work only within the minimal subalgebra $\MM$
\equ{minimal}.
Starting from this result one can obtain the BRST cohomology
with or without the fields $\ln\l$ and $\ln\bar{\l}$. The {\em local}
anomalies as well as the {\em vertex operators}, which are used to build
up the classical action cannot depend on the "nonlocal" fields $\l$
or $\bar{\l}$. Therefore the anomalies and the vertex operators are elements
of the local BRST cohomology on the field $\{ \phij,\m,\bar{\m},C,\Cb\}$.
In the next ection we shall give a proper account of this problem.

In the new fields $\{\g~,~\bar{\g}\}$ the operator $\d$ has a very
simple action
\be
\d \g =d Z~~~~~~,~~~~~~\d \bar{\g} =d\bar{Z}
\ee and we can obtain all possible {\em vertex operators}  starting from
\equ{generalsol} and applying $\f1{2}\d^2$. In this way we can obtain the
tachyon, graviton and dilaton vertex operators see \cite{lazzarini,stora}.

The tachyon vertex operator is generated by
\be
\g\bar{\g} \left[f(\PH_{0,0}(Z,\bar{Z}\right]_{Z,\bar{Z}} \label{omegaz}
\ee
where $[f]_{Z,\bar{Z}}$ is the $(Z,\bar{Z})$-componant in the "big"
indices of the function in the scalar field. This component $[f]_{Z,\bar{Z}}$
is related to the corresponding $(z,\bar{z})$-component $[f]_{z,\bar{z}}$
in the "little" indices by the relation
\be
[f]_{Z,\bar{Z}}=\f1{\l\bar{\l}}[f]_{z,\bar{z}}.
\ee
The tachyon vertex operator obtained from \equ{omegaz} has the usual form
\cite{lazzarini,stora}:
\be
(V_{z\bar{z}}(z,\bar{z}))_{tachyon}=[f(\PH(Z,\bar{Z})]_{Z,\bar{Z}}dZ\wedge
d\bar{Z}=
(1-\m\mb ) [f(\phi_{0,0})]_{z,\bar{z}}dz\wedge d\bar{z}.
\ee
The scalar function $f$ can be determined from the conformal Ward identities
governing the vertex inseration \cite{lazzarini,stora}.

The graviton and dilaton vertex operators can be obtained in the same way
from the general form of the solution for the equation \equ{fund} given by
\equ{generalsol} for differen choices possible. Therefore we can
chose the following solutions
\bea
(\o_{0})_{grav}&=&\g\bar{\g} \6_{Z}\PH_{00}(Z,\bar{Z})
\6_{\bar{Z}}\PH_{00}(Z,\bar{Z}) g(\PH_{00}(Z,\bar{Z})) \non
(\o_{0})_{dilaton}&=&\g\bar{\g} [(\6_{Z}\PH_{00}(Z,\bar{Z}))^2
h(\PH_{00}(Z,\bar{Z}))+c.c.]
\eea
where the functions $g$ and $h$ could be fixed by the conformal Ward
identities.

The corresponding vertex operators are
\bea
(V)_{grav}&=&(1-\m\mb)(D\phi_{00})(\bar{D}\phi_{00})g(\phi_{00})dz\wedge
d\bar{z} \\
(V)_{dilaton}&=&(1-\m\mb)[\f{\bar{\l}}{\l}
(D\phi_{00})^2 h(\phi_{00}) +c.c.] dz\wedge d\bar{z}.
\eea

The graviton vertex operator does not depend on $\l$ and $\lb$. This should
lead to an independence of this vertex operator from the $Z$ and $\bar{Z}$
indices. On the other hand the dilaton vertex does depend on $\l$ and
$\lb$ fact that implies the nonlocality of this vertex. In fact a
necessary but not sufficien condition for the locallity in $\m$ and $\mb$
is the disappearence of $\l$ and $\lb$ from the vertex operators
$V_{z,\bar{z}}(z,\bar{z})$ since $\l$ (and$\lb$) is a nonlocal holomorphic
function in $\m$ and $\mb$ (see \cite{lazzarini,stora,bl1}).


\section{BRST cohomology without  the variable $\ln{\l}$}
\setcounter{equation}{0}

In this section we shall calculate the BRST cohomology group for a BRST
differential algebra $\AA$ {\em without} the variables $\ln\l$ and
$\ln\bar{\l}$. The members of this cohomology are the ones which
represent the physical quantities since all operators in
a local  quantum field theory must be {\em local} field i.e. a monomial
in the basic fields and their derivatives. Besides they must be
{\em unintegrated function in the fields} which means they should be
differntial formas with coefficients analytic funtions on the local fields.

There are two ways to calculate this cohomology:
\begin{itemize}
\item To use the result of the previous section and to calculate all possible
solutions of eq.\equ{fund}, which can be even $s$-exact, and to select the
ones
without and dependence of $\l$ and $\bar{\l}$.  This is the procedure
used by Bandelloni and Lazzarini in \cite{bl1,bl2}.
\item To calculate the BRST cobomology in a {\em reduced } algebra $\AA'$,
which contains all field but $\{\l~,~\bar{\l}\}$.
\end{itemize}
We shall adopt the second point of view since in a particular case we can
reduce the present problem to the one corresponding to the BRST  cohomology
of the 2D conformal gravity. The last problem have been solved by
Brandt, Troot and Van Proeyen \cite{btp} (see alco \cite{btv} for a complete
solution ).

In the first part of this section we will accomodate the results from
\cite{btp,btv} for the Beltrami parametrization with only one matter
field $\phi_{0,0}$
and in the second part we
give some examples for the members of the BRST cohomology in the presence of
some different matter fields $\phij$.

In the case of only one matter field $\phi_{0,0}$ with the conformal
weight $(0,0)$ the minimal subalgebra $\MM$  is generated by the basis
$\{\phijp~,~C^n~,~\Cb^n\}$ and we have to calculate the solution of
the eq. \equ{fund} in this basis. All nontrivial solutions must have
the total weight (0,0). Therefore we have to eliminate many possible
solutions
and we are left with only a few possibilities.

A basis which is more convenient for our purposes is one which contains
only monomials with the total weight (0,0). Due to the fact that all
ghosts anticommute and only the ghosts $C=C^{-1}$  and $\Cb^{-1}$ have
negative weights (-1,0) respectively (0,-1) we can calculate the members
of BRST-cohomologies built up from a reduced basis
with only eight elements:
\bea
\psi_1^0=\phi_{0,0}^{0,0}\non
\psi_2^1=C^0~,~\psi_3^1=\Cb^0~,~\psi_4^1=C\phi_{0,0}^{1,0}~,~
\psi_5^1=\Cb\phi_{0,0}^{0,1} \non
\psi_6^2=C C^1~,~\psi_7^2=\Cb\Cb^1~,~\psi_8^2=C\Cb\phi_{0,0}^{0,0}.
\eea
All these elements have the total weight (0,0) and generate a  minimal
algebra since
\bea
s\psi_1^0=\psi_4^1+s\psi_5^1\non
s\psi_2^1=\psi_6^2~,~s\psi_3^1=\psi_7^2~,~s\psi_4^1=\psi_8^2~,~
s\psi_5^1=-\psi_8^2\non
s\psi_6^2=0~,~s\psi_7^2=0~,~s\psi_8^2=0.
\eea
Now by writing down all possible monomial constructed from this basis
and just by simple inspection we have found out all solutions of
eq. \equ{fund} for different values of the ghost number.
\bi
\item For ghost number g=0 and g=1 we have not found any solution;
\item for g=2 we have found only one {\em independent} solution
\be
\o^2_1=\psi^1_4 \psi^1_5 F(\psi^0_1).
\ee
where $F=F(\psi^0_1)$ is an arbitrary smooth  function of the matter field.
\item For g=3 there are four independent  solutions
\bea
\o^3_2=\psi^1_2 \psi^1_3 \psi^1_4  F(\psi^0_1) ~,~
\o^3_3=\psi^1_2 \psi^1_3 \psi^1_5 F(\psi^0_1)\non
\o^3_4=\psi^1_2 \psi^2_6  ~,~\o^3_5=\psi^1_3 \psi^2_7.
\eea
The last two solution $s$ coincide with the Guelfand and Fuks cocycles
\cite{gf}
\item For g=4 there are three independent solutions
\be
\o^4_6 =\psi^1_2\psi^1_4 \psi^2_7  F(\psi^0_1)
 ~,~ \o^4_7 =\psi^1_3\psi^1_5\psi^2_6 F(\psi^0_1).
\ee
\item For g=5 there are two independnt solutions
\be
\o^5_8 =\psi^1_2\psi^1_3\psi^1_5\psi^2_6  F(\psi^0_1)   ~,~
\o^5_9 =\psi^1_2\psi^1_3\psi^1_4\psi^2_7  F(\psi^0_1).
\ee
\item For g=6 there is only one solution
\be
\o^6_{10} =\psi^1_2\psi^1_2\psi^2_6\pi^2_7  F(\psi^0_1).
\ee
\ei
Now the member of the functional cohomology are, in fact, the solutions
of the descent equations \equ{desc} and they can be obtained using the
operator $\d$ introduced in \equ{delta} by using \equ{solut1}.
The action of $\d$ is simpler if we use the diffomorphis ghosts
$c$ and $\bar{c}$ related  to $C$ and $\Cb$ by the relations
\be
C=c+\m\bar{c}~~~,~~~\Cb=\bar{c}+\bar{\m}{c}. \ee
and the equation \equ{solut1} can be rewitten as
\be
(s+d)\o_0(c+dz,\bar{c}+d\bar{z},\Psi)=0   \label{solut5} \ee
where $\Psi$ represent all the fields except $c$ and $\bar{c}$.
Now if one use the solutions of \equ{fund} just given the
equation \equ{solut5} yields the results which are presented in
the Table .
In this Table we made use of the following notations
\bea
c^0 =\6 c +\mb\6\cb \non
\cb^0 =\pb\cb +\m\pb c \non
c^1 =\6^2 c + 2\6\mb\6\cb+\mb\6^2\cb \\
\cb^1 =\pb^2\cb +2\pb\m\pb c+\m\pb^2 c \non
y=1-\m\mb \non
\eea
\bc
\begin{tabular}{||c|c|c||} \hline\hline
$Ghost$ &     $Monomial$          &      $\delta^{2}(Monomial)/dz\wedge
d\bar{z} $      \\
\hline\hline
  $0$   &              -          &            -                \\   \hline
  $1$   &              -          &            -                \\   \hline
  $2$   &  $C\Cb\phi_{0,0}^{1,0}\phi_{0,0}^{0,1} F$     & $2(1-y)
\phi_{0,0}^{1,0}\phi_{0,0}^{0,1} F$ \\   \hline
  $3$   &  $C\Cb C^0\phi_{0,0}^{1,0}\phi_{0,0}^{0,1} F$     & $(1-y) c^0
\phi_{0,0}^{1,0}\phi_{0,0}^{0,1} F$ \\
        &   $C\Cb\Cb^0\phi_{0,0}^{1,0}\phi_{0,0}^{0,1} F$     & $(1-y)
\bar{c}^0
\phi_{0,0}^{1,0}\phi_{0,0}^{0,1} F$ \\
        &  $C C^0 C^1$          & $\6 C\6^{2}\mb - \6^{2}C \6\mb $ \\
        &   $\Cb\Cb^0\Cb^1$          & $\pb\Cb \pb^{2}\m - \pb^{2}\Cb\pb\m $
\\ \hline
  $4$   &  $C\Cb C^0\Cb^1\phi_{0,0}^{1,0} F$ &
$(1-y) \cb^0\cb^1 \phi_{0,0}^{1,0}   F $ \\
        &  $C\Cb C^0 C^1\phi_{0,0}^{0,1} F$ &
$(1-y) c^0 c^1 \phi_{0,0}^{0,1}   F $ \\
        &  $C\Cb C^0 \Cb^0\phi_{0,0}^{1,0}\phi_{0,0}^{0,1} F$ &
$(1-y) c^0 \cb^0 \phi_{0,0}^{1,0} \phi_{0,0}^{0,1}   F $ \\
\hline

  $5$   &  $C\Cb C^0\Cb^0 C^1 \phi_{0,0}^{0,1} F$ & $(1-y)c^0\cb^0 c^1
\phi_{0,0}^{0,1} F $ \\
        &   $C\Cb C^0\Cb^0\Cb^1 \phi_{0,0}^{1,0} F$ & $(1-y)c^0\cb^0 \cb^1
\phi_{0,0}^{1,0} F $ \\ \hline
  $6$   &   $C\Cb C^0\Cb^0 C^1\Cb^1 F$ & $(1-y)c^0\cb^0 c^1\cb^1 F $
 \\ \hline \hline
\end{tabular}

\vspace{1cm}
TABLE
\ec

{}From this table we can see that for a theory with only one
scalar matter field there are only ten  independent solutions of the
descent equations \equ{desc} and for ghost number bigger
then four we have no solution.

 For g=0 we get only one solution of the form
\be
(1-\m\bar{\m})\phi_{0,0}^{1,0}\phi_{0,0}^{0,1}F(\phi_{0,0})dz\wedge d\bar{z}=
(1-\m\bar{\m})D\phi_{0,0}\bar{D}\phi_{0,0}F(\phi_{0,0})dz\wedge d\bar{z}
\ee
with
$$
D=\f1{1-\m\mb}(\6-\m\pb)~~~,~~~\bar{D}=\f1{1-\m\mb}(\pb-\mb\6) $$
abd $F(\phi_{0,0})$ an analytic function of $\phi_{0,0}$. For $F=1$ we obtain
the classical action for the bosonic string in the Beltrami parametrization
\cite{b,bbs,lazzarini,stora}:
\be
 S_{cl}=\int dz d\bar{z} \f1{1-\m\mb}\left[(1+\m\mb)\6{\bf X}\cdot\pb{\bf X}
-\m\6{\bf X}\cdot\6{\bf X}-\mb\pb{\bf X}\cdot\pb{\bf X} )\right] ,
\label{classical}
\ee
with $\bX = \phi_{0,0}$.

For gh=1 we get four solutions, which can be rewitten as
\bea
\AA^1_2 &=& \int [ a_1  \m\6^2 C + a_2   \mb\pb^2 \Cb + \non &+&
\f1{1-\m\mb}(\6 c+\mb\6 \bar{c})\nabla {\bf X}\cdot \bar{\nabla}{\bf X}
F_2({\bf X}) +\non
 &+&  \f1{1-\m\mb}(\pb \bar{c}+\m\pb c)\nabla {\bf X}\cdot
\bar{\nabla}{\bf X} F_3({\bf X}) ]
 dz\wedge d\bar{z} \label{anomaly}
\eea
where
\be
\nabla= \6-\m\pb~~~~,~~~~\bar{\nabla}= \pb-\mb\6
\ee
and $a_j~,~j=1,2,$  are constants and $F_2(\bX)~,~F_3(\bX)$ are arbitrary
functions
of ${\bf X}$.
These solutions play a special role being the  possible
candidates for anomalies. Actually the matter dependent part of the anomaly
\equ{anomaly} cannot be associated to a true diffeomopphism anomaly if
one use as a classical action \equ{classical} since this sction does not
contain a self-interaction term in the matter fields. It follows that
in the framework of the perturbation theory the numerical coefficients of
the corresponding Feynman diagrams automatically vanishes i.e. in this
case  $a_3=a_4=0$ and the unique breaking of the diffeomorphism invariance
at the quantum level has the form
\be
\AA^1_2 = \int \left[ a_1  \m \6^2 C + a_2   \mb\pb^2 \Cb \right ]
 dz\wedge d\bar{z}. \label{s1}
\ee

For gh=2 there are three independent solutions of the following form
\bea
\AA^2_2 & = & \int [(\pb \cb + \m \pb c)(\pb^2\cb +2\pb\m\pb c+\m\pb^2 c )
\nabla {\bf X}F_4(\bX) \non
& + & (\6 c +\mb \6 \cb )(\6^2 c + 2 \6 \mb \6 \cb + \mb\6^2\cb)
{\bar{\nabla}}\bX F_5(\bX) \non
& + & (\6 c +\mb\6\cb )
(\pb\cb +\m\pb c )\nabla \bX \bar{\nabla} \bX F_6(\bX)]
\label{s2}
\eea
where $F_4(\bX)~,~F_5(\bX),~F_6(\bX)$ are arbitrary functions of $\bX$.

For gh=3 and gh=4 the solutions in number of two, respectively one,
can be rewritten as
\bea
\AA^3_2 &=& \int
[(\6 c +\mb\6\cb )
(\pb\cb +\m\pb c)
(\6^2 c + 2\6\mb\6\cb+\mb\6^2\cb ){\bar{\nabla}}\bX F_7(\bX) \non
&+&(\6 c +\mb\6\cb )
(\pb\cb +\m\pb c)
(\pb^2\cb +2\pb\m\pb c+\m\pb^2 c)
\nabla {\bf X}F_8(\bX)]
\label{s3}
\eea
and
\be
\AA^4_2 = \int [\f{1}{1-\m \mb }
(\6 c +\mb\6\cb )
(\pb\cb +\m\pb c) \non
(\6^2 c + 2\6\mb\6\cb+\mb\6^2\cb )
(\pb^2\cb +2\pb\m\pb c+\m\pb^2 c)
\nabla \bX \bar{\nabla} \bX F_9(\bX)]
\label{s4}
\ee
where $F_7(\bX)~,~F_8(\bX),~F_9(\bX)$ are arbitrary functions of $\bX$.

In the situation when there are more than one matter field
$\{ \phi_{j_{1},{\bar{j}}_{1}} \cdots \phi_{j_{n},{\bar{j}}_{n}} \}$
with the conformal
weights $ (j_{1},\bar{j}_{1})\cdots (j_{n},\bar{j}_{n})$ the simplicity of
the
previous basis dissapears since in this case there are an infinite number of
possibilities to construct local functions with the total weight $(0,0)$.

For gh=2 we take the solutions of the \equ{fund} of the form
\be
\o_0=C\Cb \phiju\cdots\phijn=C\Cb \Pi. \label{solution}
\ee
Since the total weight of \equ{solution} must be (0,0) we have to
impose the following conditions on the indices
\bea
j_1+\cdots +j_n+p_1+\cdots p_n&=&1\non
\bar{j}_1+\cdots +\bar{j}_n+q_1+\cdots q_n&=&1
\eea
The equations \equ{fund} and \equ{brst8} yield
\be
s\omega _0^2 = c\cb [-c^0 -\cb^0 + \sum _{k=-1}(c^k L_k
+ \cb ^k \bar{L}_k )]\Pi = 0 \label{equation}
\ee
since $L_k $ and $\bar{L}_k $ are even derivatives. Taking into account
that
\bea
L_k \phi^{p,q}_{j,\bar{j}} & = & 0\hspace{0.5cm}
for\hspace{0.5cm} k>p~~~or~~~k=p,j=0 \non
\bar{L}_k \phi^{p,q}_{j,\bar{j}} & = & 0\hspace{0.5cm}
for\hspace{0.5cm} k>q~~~or~~~k=q,\bar{j}=0
\eea
equation \equ{equation} yields
\bea
p_l &=&0 ~~~or ~~~p_l =1~,~j_l = 0 \non
q_l &=&0 ~~~or ~~~q_l =1~,~\bar{j}_l =0
\eea
With these conditions at hand we can write down
the most general form of solution of the equation \equ{fund}
\be
\phi^{0,0}_{j_{1},\bar{j}_1} \cdots \phi^{0,0}_{j_{n},\bar{j}_n}
\phi^{1,0}_{0,\bar{k}_1} \cdots \phi^{1,0}_{0,\bar{k}_m}
\phi^{0,1}_{l_{1},0} \cdots \phi^{0,1}_{l_{s},0}
(\phi^{1,1}_{0,0})^t
\ee
with
\be
j_1 + \cdots j_n + \bar{j}_1 + \cdots \bar{j}_n +
\bar{k}_1 + 1 + \cdots \bar{k}_n + 1+ l_1 + 1 +\cdots l_n + 1 +
2q = 1
\ee

\section{Conclusions}
\setcounter{equation}{0}
We have calculated the complete BRST  cohomology in the space of the local
functions for the local field theories on a Riemann surface which contain
the conformal matter field coupled with a complex structure parametrized
by a Beltrami differential without any reference to metrics.

For the theory with only one scalar matter field with  $\phi _{0,0}$ and
{\em without } the integrating factor $\lambda ~, ~\bar{\lambda}$, we have
calculated {\em all } members of the BRST cohomology.
For the theory with several matter fields we have found  out only a
limited number of $H(s)$, but they include all cases  considered by
Bandelloni and Lazzarini \cite{bl1,bl2}.

The simplest case is the one where
we have introduced the integrating factors $\l $ and $\bar{\l }$. Here the
BRST cohomology contains only terms of the form
\be
C\Cb\PHjpu \cdots \PHjpn \label{general}
\ee
i.e., $H^g (s)=0$ if the ghost number g$\neq $2. However, it is worth
reminding
the reader that the form \equ{general} has been
obtained only from geometrical point of view. If we want to impose the
{\em locality } asumption for our model then the factors $\l $, $\bar{\l }$
should disapear in \equ{general} since  $\l $ is a {\em nonlocal} holomorphic
function.
Thus the locality assumption bounds considerabily the
form of the solutions of \equ{fund}. We could assure the locality from
the very begining if we start to calculate the BRST cohomology without
the fields $\l $, $\bar{\l }$.

The technique presented in this paper can be used to study the BRST
cohomology for
other models, as $W_3$ - gravity \cite{tb} or the superstring
in the super-Beltrami parametrization \cite{tv}. Also it has been used to
calculate the BRST cohomology of the Slavnov operator \cite{btp} or the
BRST-antibracket cohomology for 2D gravity.

\end{document}